# Efficient bifurcation and parameterization of multi-dimensional combustion manifolds using deep mixture of experts: an *a priori* study

Opeoluwa Owoyele[*], Prithwish Kundu, Pinaki Pal

Energy Systems Division, Argonne National Laboratory, Lemont, IL 60439, USA

*Corresponding author. Email: oowoyele@anl.gov

___

**ABSTRACT**

This work describes and validates an approach for autonomously bifurcating turbulent combustion manifolds to divide regression tasks amongst specialized artificial neural networks (ANNs). This approach relies on the mixture of experts (MoE) framework, where each neural network is trained to be specialized in a given portion of the input space. The assignment of different input regions to the experts is determined by a gating network, which is a neural network classifier. In some previous studies [1-4], it has been demonstrated that bifurcation of a complex combustion manifold and fitting different ANNs for each part leads to better fits or faster inference speeds. However, the manner of bifurcation in these studies was based on heuristic approaches or clustering techniques. In contrast, the proposed technique enables automatic bifurcation using non-linear planes in high-dimensional turbulent combustion manifolds that are often associated with complex behavior due to different dominating physics in various zones. The proposed concept is validated using 4-dimensional (4D) and 5D flamelet tables, showing that the errors obtained with a given network size, or conversely the network size required to achieve a given accuracy, is considerably reduced. The effect of the number of experts on inference speed is also investigated, showing that by increasing the number of experts from 1 to 8, the inference time can be approximately reduced by a factor of two. Moreover, it is shown that the MoE approach divides the input manifold in a physically intuitive manner, suggesting that the MoE framework can elucidate high-dimensional datasets in a physically meaningful way.

**Keywords:** *flamelet modeling, multi-dimensional manifold, deep learning, mixture of experts*

___

# 1. Introduction

Numerical simulations of turbulent combustion with direct integration of detailed chemical kinetics in unsteady three-dimensional (3D) flows are computationally prohibitive, as they require solving transport equations for a large number of species in space and time [5]. One promising method to circumvent this issue is flamelet-based modeling, where the thermochemical state space, considered to be confined to a lower-dimensional manifold [6], is precomputed (based on relevant canonical flame configurations) and stored as functions of a reduced set of reaction coordinates in tables, which are then used for lookup during computational fluid dynamics (CFD) simulations of turbulent combustion. This offers significant computational savings, since transport equations are solved for a few reaction coordinates rather than all the species. However, the size of the lookup table scales exponentially with the number of dimensions, which rapidly increases the memory requirement to store these tables. In addition, the time spent on interpolation from the lookup table also increases with the size of the manifold, adversely affecting the simulation speed. Hence, the issue of dimensionality is important to address for advancing the application of tabulated flamelet models.

In previous studies, artificial neural networks (ANNs) have been used to predict or represent chemical kinetics [7-10], provide closure in various contexts [4, 11], and learn flamelet tables [12]. In a previous study by the authors [13], a framework for assimilating flamelet tables using function-approximating ANNs with lumped species in the output layers, was introduced and demonstrated in simulations of engine combustion network (ECN) spray A and a compression-ignition engine. It was found that bifurcating the flamelet manifold, where separate ANNs are used for fitting different parts of the manifold, was beneficial. This was done by "heuristically" splitting the manifold based on the scalar dissipation rate. Other studies also found that splitting the input space is beneficial [1, 2, 7], as the split manifolds have less complex surfaces compared to the entire surface. One common thread among all these previous works is that the bifurcation of the manifolds was performed by splitting the input space without a rigorous framework that takes the functional form of the problem into consideration. For example, self-organizing maps [1] are based on proximity in input space, and therefore, often inefficient since proximity does not necessarily translate to similarity in physical behavior. On the other hand, splitting based on a given independent variable is based on trial-and-error. Moreover, simply bifurcating in one input variable may not adequately reduce the complexity that exists due to the effects of other variables.

In this context, this work presents a Mixture of Experts (MoE) [14, 15] technique for representing flamelet manifolds. In this approach, the clustering process is merged with regression, where a classifier is trained as a gating network in conjunction with a number of neural network regressors called "experts".

These experts compete for samples during training, and this results in a bifurcated manifold, where different networks are experts in predicting the target variable of interest in different regions of the manifold, depending on the behavior and magnitude of the target variable. There are two major benefits to this approach. The first is that smaller networks can be used, resulting in much faster inference speeds compared to one large network for the entire surface, since fewer floating-point operations (FLOPs) would be needed to retrieve the target variable. A second benefit is that lower errors can be obtained if the manifold is bifurcated while the size of the networks remains constant. Overall, this method can automatically bifurcate a multi-dimensional manifold *based on its functional form* and train a collection of neural networks that accurately represent the manifold at lower retrieval costs with negligible memory footprint.

The rest of the paper is organized as follows. In section 2, the basic formulation of the MoE approach for assimilating flamelet tables is presented along with a brief description of the test problem chosen for *a priori* demonstration of the technique. Main results associated with the accuracy and inference speed of the new approach and physical relevance of the manifold bifurcation are discussed in the subsequent section. The paper finishes with concluding remarks in section 4.

## 2. MoE: Theory and mathematical formulation

The MoE approach [14, 15] (Fig. 1) trains multiple feed-forward ANN regressors known as experts, where each expert is specialized to handle a specific region of the input space. The assignment of these regions is determined by another ANN, called a gating network, which is typically a feed-forward classifier. During training, the gating network and all the experts are trained simultaneously via the backpropagation of errors.

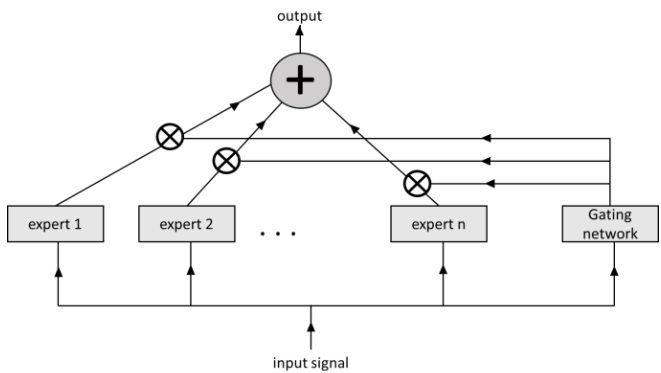

**Fig. 1.** A schematic of the MoE framework.

Increasingly complex functions require neural networks with more free parameters for fitting. As stated in the previous section, such a complex surface can be divided, and different parts can subsequently be allotted to appropriate experts for fitting. As a result, the function that each expert fits is less complex, and thus accuracy can be significantly improved. The prediction from a mixture of *N* experts is given by:

$$\Phi = \sum_{k=1}^{N} p_k \varphi_k \tag{1}$$

where, $\varphi_k$ is the output of expert *k,* while $p_k$ is the output of the gating network and represents the prior probability that expert *k* can produce the best prediction based on the current sample. Ideally, the gating network would give a probability of 1 for the appropriate expert and probability of 0 for others. In this way, the gating network would provide a criterion for conditionally performing inference with the experts, because the prediction of only *one* network needs to be computed for a given input signal. However, this does not always occur since the gating network predicts intermediate (between 0 and 1) values at the boundary of two experts. Therefore, in this work, conditional computation is enforced via a second training step, where experts are exclusively trained on areas of the manifold where they are the best experts. In this way, the regions are divided among the experts with hard boundaries in a non-overlapping mode.

In this work, as introduced by Jacobs et al. [15], the experts are treated as a mixture of Gaussians. The loss, *L*, for *k* experts is defined as follows:

$$L = \frac{-log \sum_{k=1}^{N} p_k \exp(-\beta \|e_k\|^2)}{2\beta} \tag{2}$$

In Eq. (2), $e_k = Y - \varphi_k$, where *Y* is the desired output as obtained from the flamelet table and $e_k$ is the corresponding error. The derivative of *L* with respect to the output of expert *k* is the following.

$$\frac{dL}{d\varphi_k} = \alpha_k \|e_k\| \tag{3}$$

where

$$\alpha_k = \frac{p_k \exp\left(-\frac{\beta}{2}\|e_k\|^2\right)}{\sum_{k=1}^{N} p_k \exp\left(-\frac{\beta}{2}\|e_k\|^2\right)} \tag{4}$$

The weighting terms ($\alpha_k$'s) encourage specialization among the experts, since $\alpha_k$ is large when $\|e_k\|$ is small and vice versa. One way to interpret $\alpha_k$ from Eq. (4) is that it gives the posterior probability that an expert $k$ is the best expert to produce the desired output, $Y$.

It must be noted that the parameter, $\beta$, in Eq. (2) represents a refinement added to the basic MoE approach [15] to improve performance. It results in the squared-norm of the error being scaled by a parameter $\beta$, to promote scale-invariant separation of the experts. In order to show why this is important, consider a simplified case without $\beta$ in Eq. (4) where there are two experts which have equal prior probabilities from the gating network. If $\|e_k\| = 1$ for expert 1, and $\|e_k\| = 10$ for expert 2, based on Eq. (4), $\alpha_1 \approx 1$, and $\alpha_2 \approx 0$, showing a clear separation between the two experts in this case. However, if $\|e_k\| = 0.01$ for expert 1, and $\|e_k\| = 0.1$ for expert 2, $\alpha_1 \approx 0.5012$, and $\alpha_2 \approx 0.4988$. In other words, as the magnitudes of the errors decrease, the separation between the experts becomes more ambiguous, even when the ratio of the errors remains the same. The introduction of $\beta$ circumvents this issue by scaling the errors, such that the two experts remain separated even as the magnitudes of the errors decrease. To achieve this, $\beta$ is dynamically adjusted during training such that each expert maintains a minimum posterior probability of 0.99. For a two-expert mixture with equal prior probabilities, $p$, it can be derived from Eq. (4) that $\beta$ is given by:

$$\beta = \frac{\log \alpha_1 - \log(1 - \alpha_1)}{\|e_1\|^2 - \|e_2\|^2} \tag{5}$$

In practice, $\beta$ is calculated based on the weakest expert at its worst-performing sample over the entire dataset, setting $\alpha_1$ to 0.99 in Eq. (4). Apart from keeping both experts specialized, this ensures that $\beta$ is adjusted such that the weaker expert always has a minimum posterior probability of 0.99. Thus, during training, it helps to keep the weaker expert from being overpowered by the stronger expert.

In this work, a hierarchical approach is employed, where the top-level gating network bifurcates the manifold between two experts, with each expert being a MoE. This recursive splitting continues until the desired number of experts and accuracy is achieved. In the beginning, the networks are all initialized randomly. As training progresses, the input space gets neatly divided among the networks. During a given training epoch, whenever an expert produces lower errors for a given sample, the gating network's weights are adjusted such that it recognizes that the sample should be assigned to this expert. Thus, experts that fit specific samples better are in the future rewarded with stronger signals for training these samples, and this promotes specialization among the networks. On the other hand, experts that perform poorly are progressively presented with weaker signals but are encouraged with stronger signals in other regions of input space where they function well.

The above MoE framework is demonstrated in the context of the unsteady flamelet progress variable (UFPV) model [16, 17], where the mass fraction of species, *i*, is parameterized as a function of mixture fraction (*Z*), mixture fraction variance ($Z_v$) progress variable (*C*), and scalar dissipation rate ($\chi$) as:

$$Y_i = Y_i(Z, Z_v, \chi, C) \tag{6}$$

The flamelet equations were solved *a priori* based on an unsteady, auto-igniting, n-dodecane counter-flow diffusion flame configuration, under a pressure of 60 bar and oxidizer temperature of 900 K. A 103 species skeletal mechanism [18] was used to solve the unsteady flamelet equations using an in-house code. The flamelet solutions were tabulated as a four-dimensional (4D) manifold. This manifold with complex chemical kinetics is used in this study to demonstrate and validate the MoE approach.

## 3. Results and discussion

### *3.1 A-priori validation*

In the hierarchical approach employed in this work, a 3-level MoE framework was implemented in an in-house python code built using Tensorflow [19]. The manifold was recursively split into two until a total of 8 experts were obtained. The gating network was a single-layer feed-forward neural network with a sigmoid activation [20] in the hidden layer and standard softmax activation in the output layer. Each expert was a 4-layer deep ANN with 12 neurons activated by the leaky rectified linear unit (Leaky ReLU) [21] transfer function in each hidden layer. Henceforth, for simplicity, the ANN architecture is defined as (number of neurons × number of layers), for example, 12 × 4 for the above network architecture. In this subsection, the accuracy of the MoE approach is compared against the conventional approach of fitting the entire manifold using a single ANN (SANN).

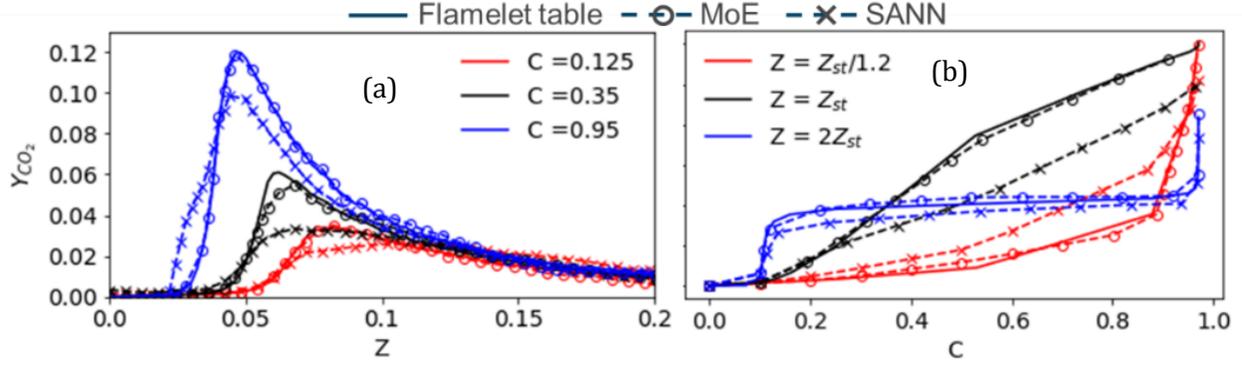

**Fig. 2.** $Y_{CO_2}$ versus $Z$ (left) and $C$ (right) as obtained from the flamelet table, MoE and SANN.

In Figs. 2 and 3, one-dimensional plots of mass fractions of a few selected species versus various independent variables are shown. In both figures, the slices shown are at $Z_v = 0.0$ and $\chi = 5$ s$^{-1}$, unless stated otherwise. These slices are selected from a wide range of possible parametric combinations in 4D space the sake of brevity. Here, $12 \times 4$ ANNs with Leaky ReLU activation function in the hidden layers were used for the MoE approach.

Figure 2a shows the mass fraction of $CO_2$ ($Y_{CO_2}$) as a function of mixture fraction, $Z$, for different levels of $C$. It is evident that the profile of $Y_{CO_2}$ is initially flat, then rises sharply and peaks before slowly tapering down as $Z$ tends towards 1. For higher values of $C$, this peak occurs around the stoichiometric value of $Z$ ($Z_{st} \approx 0.05$), while a shift of the peak towards richer conditions occurs for lower values of $C$. For all levels of $C$, the profiles start from the same origin where the mass fractions are *0* at *Z = 0*, and then diverge as $Z$ increases, before converging again as the value of $CO_2$ mass fraction diminishes, without the lines intersecting. This behavior is well captured by the MoE. On the other hand, the predictions from SANN shows significantly higher errors, with the profiles for different levels of $C$ sometimes intersecting, and the peaks being significantly underpredicted. Figure 2b shows $Y_{CO_2}$ against $C$ for $Z_{st}$, $2Z_{st}$, and $Z_{st}/1.2$. For $Z = Z_{st}$, it is almost a linear function of $C$, albeit with some subtle irregularities. There is a flat region early on ($C < 0.1$), followed by a region where the mass fraction rises linearly ($0.2 < C < 0.5$), and then another linear region with a slightly lower slope ($C > 0.5$). Again, the MoE picks up the slight change in slope at $C \approx 0.5$, while only slightly underpredicting $Y_{CO_2}$ around this region. The SANN, on the other hand, simply models this as a straight line, for the most part. Besides, for the other $Z$ levels shown ($Z_{st}/1.2$ and $2Z_{st}$) as well, the profiles are captured by the MoE with much higher accuracy.

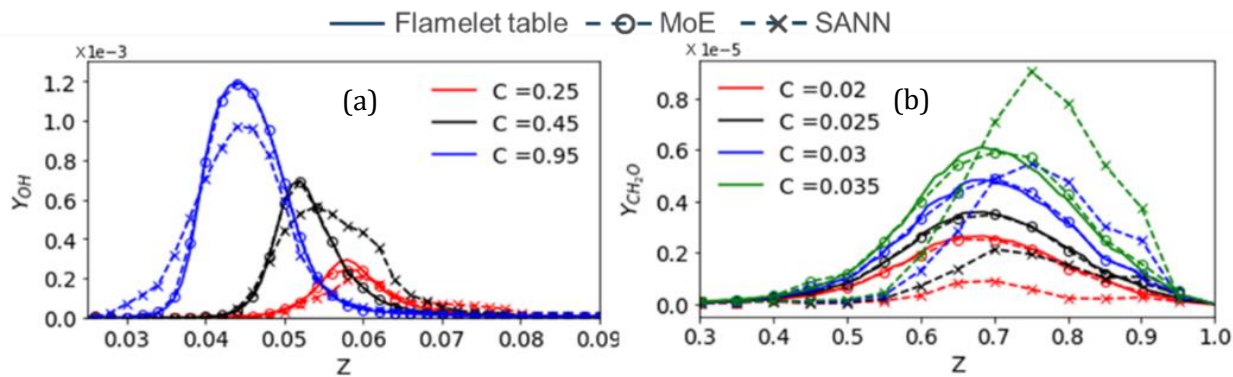

**Fig. 3.** $Y_{OH}$ (left) and $Y_{CH2O}$ (right) versus $Z$ as obtained from the flamelet table, MoE and SANN.

Similar plots for OH and CH$_2$O are shown in Fig. 3. In Fig. 3a, the line plots of $Y_{OH}$ as a function of $Z$ show trends similar to those obtained for CO$_2$, with peaks occurring at slightly leaner regions for higher levels of $C$ and peaks shifted towards the richer zones as $C$ reduces. There are, however, two differences. First, the regions of increased $Y_{OH}$ occur in significantly narrower bands in $Z$ space, with a sharp decline as we move away from $Z_{st}$. Secondly, as opposed to CO$_2$ where the profiles for different levels of $C$ do not intersect, the relative shift in peaks is severe enough so that the profiles for different levels of $C$ intersect. This behavior is captured by the MoE, while once more, significantly larger errors are observed for SANN. In general, the SANN underpredicts the peaks similar to CO$_2$, and in some cases, the profile shapes are not well captured (for instance, at $C = 0.45$). CH$_2$O is an intermediate species and plays an important role in the low temperature combustion regime. The profiles of $Y_{CH2O}$ are shown in Fig. 3b. The slices shown here are at a higher scalar dissipation rate of 300 s$^{-1}$, since the peaks of CH$_2$O concentration occur at high scalar dissipation rates where the flame is beyond the ignition limit. It can be observed that the peak CH$_2$O mass fractions occur at much richer regions in the mixture fraction space, compared to CO$_2$ and OH. In addition, a shift in the peaks is observed with reducing $C$, but in the opposite direction to that of OH. For the network size used here, SANN struggles to produce reasonable fits for all levels of $C$, unlike MoE.

The bar graphs in Fig. 4 summarize the mean squared errors (MAEs) for MoE with different number of neurons per hidden layer and number of experts; only two species, CO$_2$ and OH, are shown for the sake of brevity. The MAEs are based on the best case result out of 10 independently trained networks, each with a different random re-initialization of weights. "One expert" in the figure corresponds to SANN. Expectedly, for a given number of experts, the error drops as the number of neurons per hidden layer is increased. It can also be seen that for a given network architecture, the error decreases as the number of experts increases. However, there are two drawbacks to increasing the size of the neural network: (a) it increases the dimensionality of the weights that are optimized during training, thereby rendering the networks more difficult to train, and (b) number of FLOPs needed to retrieve the chemical states of the system when these networks are coupled to CFD solvers increases, thereby slowing down inference and hence, the CFD simulation. On the other hand, splitting up the manifold and using multiple smaller neural networks increases both convergence rates during training and inference speeds. In order to demonstrate this, a standalone code was developed in FORTRAN to compare the inference speed of a $4 \times 4$ network and a $12 \times 4$ network. Without considering the inference time of the gating network, a $4 \times 4$ neural network was found to be approximately four times faster than a $12 \times 4$ network, independent of the number of experts. After considering the time taken to run the gating networks, the inference speed-up was still around a factor of 2. It should be noted, however, that the additional time to run the gating network is constant, irrespective of the size of the experts. In this case, the gating network takes approximately 50% of inference time. For more complex problems with larger networks, higher speedups would be expected, since the gating network would have a less significant effect on the total inference time.

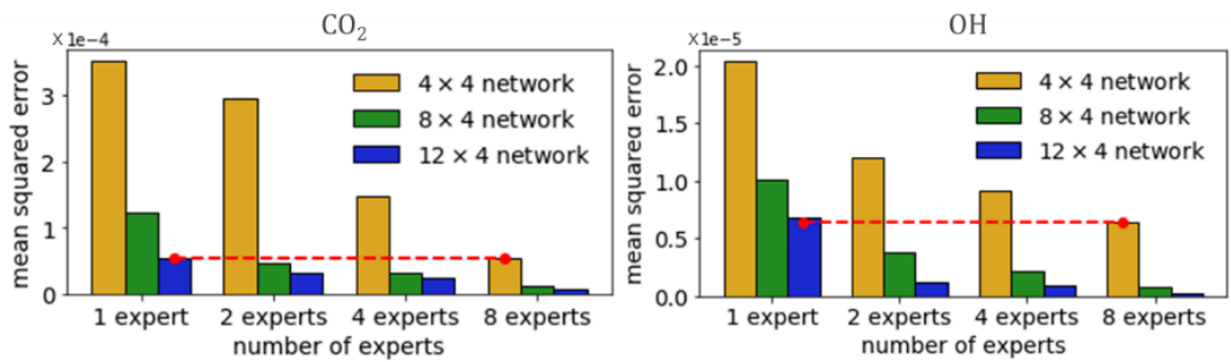

**Fig. 4:** MAEs obtained with MoE using various network configurations and different number of experts.

As evident from Fig. 4, the MoE approach offers much higher accuracy than SANN. For example, for $CO_2$, with the baseline case of a $12 \times 4$ network, the mean squared errors are reduced by an order of magnitude when MoE with 8 experts is used as opposed to SANN. On the other hand, assuming the accuracy of the baseline configuration to be sufficient, inference speed can be further improved by using more experts with smaller networks. For example, it can be seen that similar errors to the baseline case can be obtained by using a mixture of eight $4 \times 4$ experts (illustrated by the red dotted lines).

To further analyze the efficacy of the MoE approach, similar studies were conducted for a five-dimensional (5D) flamelet manifold, where 9 levels of pressure were added to the 4D manifold described in section 2. Again, significantly improved accuracy was achieved (similar to the 4D case) with superior inference speeds compared to SANN, as shown in Fig. 5. Notably, the 5D flamelet table had approximately 50 million samples, resulting in a file size of about 14 gigabytes, which would introduce significant storage requirement issues on compute clusters, if used in a CFD simulation. The use of MoE to efficiently parametrize such large manifolds drastically brings down this memory footprint to mere kilobytes, which would enable simulations of practical combustion systems (such as, internal combustion engines) with flamelet models.

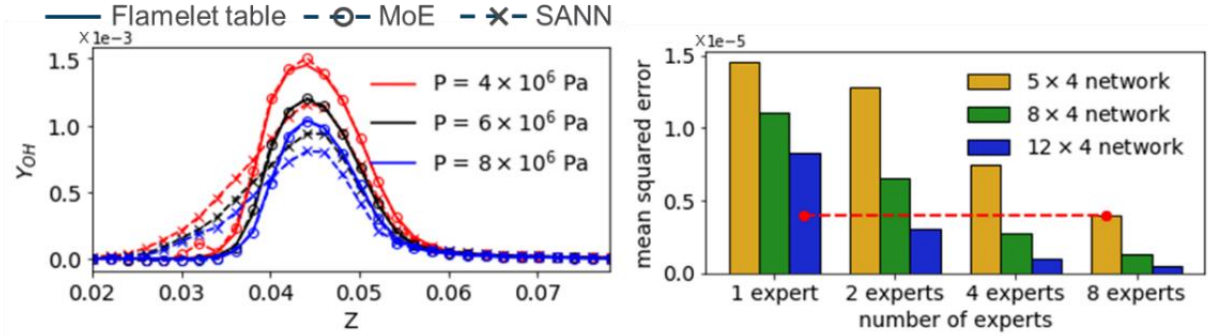

**Fig. 5:** $Y_{OH}$ versus $Z$ and MAE for $Y_{OH}$ predictions for a 5D table.

### 3.2 Physical relevance of the bifurcations

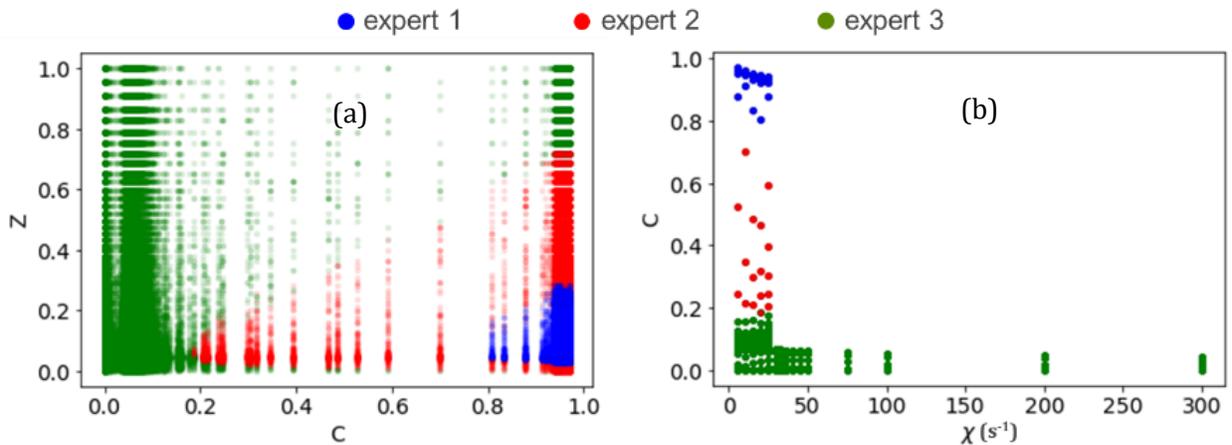

**Fig. 6.** Regions of the flamelet manifold assigned to various experts for $Y_{CO2}$ training.

In this section, the significance of the partitions obtained from the MoE is briefly discussed. Here, the case of MoE consisting of a single level (no recursive splitting) with 3 experts is chosen for ease of interpretation and visualization, since manifold partitions may be more difficult to understand for larger number of experts. As an example, we consider the splitting of the manifold for the prediction of $Y_{CO2}$. 2D scatter plots of $Z$ versus $C$ and $C$ versus $\chi$ are depicted in Fig. 6. From Fig. 6a, it can be seen that at $C \approx 1$, samples around $Z_{st} \approx 0.05$ are assigned to expert 1. The region occupied by expert 1 narrows as $C$ tends to 0, finally cutting off at $0.7 < C < 0.8$. This region, assigned to expert 1, roughly corresponds to rapid rise and high $CO_2$ concentrations. Expert 2 connects the region between those that expert 3 and expert 1 specialize in. Further visualization in Fig. 6b suggests that expert 1 is predominantly assigned post-ignition zones, expert 3 is mostly assigned pre-ignition zones, while expert 2 is assigned the ignition zones. Experts 1 and 2 are exclusively assigned points where $\chi < 25$ s$^{-1}$, since the flamelets don't ignite above $\chi = 25$ s$^{-1}$ in this case.

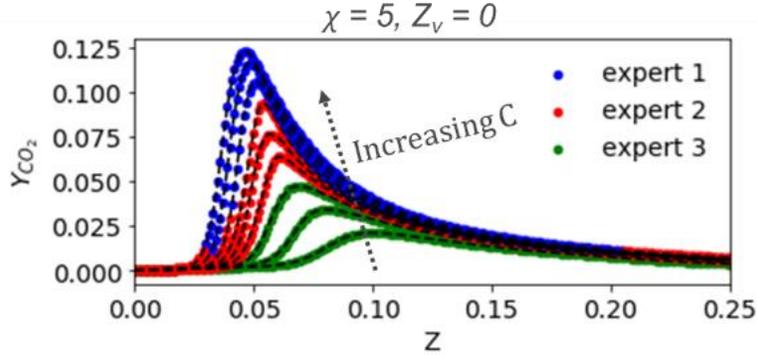

**Fig. 7.** 1D profiles of $CO_2$ mass fraction showing the assignment of various sections to different experts.

Another depiction of the manifold partition is provided in Fig. 7, showing 1D plots of $Y_{CO2}$ versus $Z$ for different levels of $C$. It can be seen that for higher levels of $C$, the regions around the peaks are assigned to expert 1, while expert 2 is assigned the flatter regions. As $C$ decreases, increasingly narrower zones close the peaks are assigned to expert 1, until the slopes and peaks of $Y_{CO2}$ are too low causing the regions close to the peaks to be assigned to expert 2. At even lower values of C, the slopes and values of $Y_{CO2}$ do not rise high enough, and all the points are assigned to expert 3 for all values of $Z$.

## 4. Conclusions

In this work, a divide-and-conquer framework was developed to efficiently parameterize complex turbulent combustion manifolds. The approach was based on a MoE framework, where a collection of deep neural networks is trained with different networks specialized in different portions of the manifold. The proposed approach was tested on 4D and 5D flamelet tables, showing significant benefits in terms of accuracy, inference speed, and storage requirement. A discussion of the physical significance of bifurcations was also presented, showing that the MoE approach divides the manifold in a physically meaningful way. By dividing the regression tasks based on physical behavior, this approach helped to elucidate the manifold of interest. While the MoE approach was applied to flamelet manifolds in this work, this method can be seamlessly extended to other areas in turbulent combustion modeling, such as tabulation of the source terms of principal components [3, 11], prediction of chemical source terms for subgrid models [10], and prediction of pyrolysis of complex fuels [9].

In future studies, the proposed algorithm will be coupled with CFD solvers to perform *a posteriori* validation studies for canonical turbulent flames as well as practical internal combustion and gas turbine engines. While in the present work, the MoE approach has been used for parametrizing 4D/5D tables, more significant benefits can be reaped by parameterizing higher-dimensional manifolds where the required memory is above the available memory per node. Furthermore, with hybrid CPU/GPU configurations in high-performance computing systems, techniques like online in-situ training, or GPU-based machine learning inference can be exploited for even faster inference with MoE on these future platforms. In this way, the use of deep learning to assimilate multi-dimensional manifolds can provide huge benefits, as it will enable the addition of more independent variables to capture effects that would otherwise be impossible due to the inefficient scaling of traditional interpolation-based methods.

## Acknowledgements

The submitted manuscript has been created by UChicago Argonne, LLC, Operator of Argonne National Laboratory (Argonne). Argonne, a U.S. Department of Energy Office of Science laboratory, is operated under Contract No. DE-AC02-06CH11357. The U.S. Government retains for itself, and others acting on its behalf, a paid-up nonexclusive, irrevocable worldwide license in said article to re- produce, prepare derivative works, distribute copies to the public, and perform publicly and display publicly, by or on behalf of the Government. The research at Argonne National Laboratory was funded by DOE's Office of Vehicle Technologies, Office of Energy Efficiency and Renewable Energy. The authors gratefully acknowledge the computing resources provided on Blues, a high-performance computing cluster operated by the Laboratory Computing Resource Center at Argonne National Laboratory.

# References

[1] J. Blasco, N. Fueyo, C. Dopazo, J. Chen, A self-organizing-map approach to chemistry representation in combustion applications, Combust. Theor. Modell., 4 (2000) 61-76.

[2] L.L. Franke, A.K. Chatzopoulos, S. Rigopoulos, Tabulation of combustion chemistry via Artificial Neural Networks (ANNs): Methodology and application to LES-PDF simulation of Sydney flame L, Combust. Flame 185 (2017) 245-260.

[3] O. Owoyele, T. Echekki, Toward computationally efficient combustion DNS with complex fuels via principal component transport, Combust. Theor. Modell., (2017) 1-29.

[4] R. Ranade, T. Echekki, A framework for data-based turbulent combustion closure: A posteriori validation, Combust. Flame 210 (2019) 279-291.

[5] S.B. Pope, Small scales, many species and the manifold challenges of turbulent combustion, Proc. Combust. Inst. , 34 (2013) 1-31.

[6] N. Peters, Turbulent combustion, Cambridge University Press, 2000.

[7] J. Blasco, N. Fueyo, J. Larroya, C. Dopazo, Y.-J. Chen, A single-step time-integrator of a methane–air chemical system using artificial neural networks, Comput. Chem. Eng., 23 (1999) 1127-1133.

[8] F. Christo, A. Masri, E. Nebot, S. Pope, An integrated PDF/neural network approach for simulating turbulent reacting systems, Proc. Combust. Inst., 26 (1996) 43-48.

[9] R. Ranade, S. Alqahtani, A. Farooq, T. Echekki, An ANN based hybrid chemistry framework for complex fuels, Fuel, 241 (2019) 625-636.

[10] B.A. Sen, S. Menon, Linear eddy mixing based tabulation and artificial neural networks for large eddy simulations of turbulent flames, Combust. Flame 157 (2010) 62-74.

[11] T. Echekki, H. Mirgolbabaei, Principal component transport in turbulent combustion: A posteriori analysis, Combust. Flame 162 (2015) 1919-1933.

[12] M. Ihme, C. Schmitt, H. Pitsch, Optimal artificial neural networks and tabulation methods for chemistry representation in LES of a bluff-body swirl-stabilized flame, Proc. Combust. Inst. , 32 (2009) 1527-1535.

[13] O. Owoyele, P. Kundu, M.M. Ameen, T. Echekki, S. Som, Application of deep artificial neural networks to multi-dimensional flamelet libraries and spray flames, Int. J. Engine Res., (2019) 1468087419837770.

[14] R.A. Jacobs, M.I. Jordan, A.G. Barto, Task decomposition through competition in a modular connectionist architecture: The what and where vision tasks, Cogn. Sci., 15 (1991) 219-250.

[15] R.A. Jacobs, M.I. Jordan, S.J. Nowlan, G.E. Hinton, Adaptive mixtures of local experts, Neural comput., 3 (1991) 79-87.

[16] C. Bajaj, M. Ameen, J.J.C.S. Abraham, Evaluation of an unsteady flamelet progress variable model for autoignition and flame lift-off in diesel jets, Combust. Sci. Technol., 185 (2013) 454-472.

[17] M. Ihme, Y.C.J.C. See, Prediction of autoignition in a lifted methane/air flame using an unsteady flamelet/progress variable model, Combust. Flame, 157 (2010) 1850-1862.

[18] Z. Luo, S. Som, S.M. Sarathy, M. Plomer, W.J. Pitz, D.E. Longman, T. Lu, Development and validation of an n-dodecane skeletal mechanism for spray combustion applications, Combust. Theor. Modell., 18 (2014) 187-203.

[19] M. Abadi, A. Agarwal, P. Barham, E. Brevdo, Z. Chen, C. Citro, G.S. Corrado, A. Davis, J. Dean, M. Devin, Tensorflow: Large-scale machine learning on heterogeneous distributed systems, arXiv preprint arXiv:.04467, (2016).

[20] P.-F. Verhulst, Mathematical researches into the law of population growth increase, J Nouveaux Mémoires de l'Académie Royale des Sciences et Belles-Lettres de Bruxelles, 18 (1845) 1-42.

[21] A.L. Maas, A.Y. Hannun, A.Y. Ng, in: Proc. icml, 2013, pp. 3.